\documentclass[fleqn,12pt,twoside]{article}
\usepackage{espcrc1}

\usepackage{graphicx}

\newcommand{\AmS}{{\protect\the\textfont2
  A\kern-.1667em\lower.5ex\hbox{M}\kern-.125emS}}
\title{Universality in the physics of cold atoms with large scattering
 length\footnote{
This work was done in collaboration with Eric Braaten.
}}

\author{H.-W. Hammer\address[HISKP]{
        Universit\"at Bonn, 
        HISKP (Theorie),
        Nussallee 14-16, 53115 Bonn, Germany}}%
       
\begin{document}

\maketitle

\begin{abstract}
Effective field theories exploit a separation of scales in physical 
systems in order to perform systematically improvable, model-independent 
calculations. They are ideally suited to describe universal aspects
of a wide range of physical systems. I will discuss recent applications
of effective field theory to cold atomic and molecular few-body
systems with large scattering length.
\end{abstract}

\section{INTRODUCTION}
The Effective Field Theory (EFT) approach provides a powerful framework 
that exploits the separation of scales in physical systems.
Only low-energy (or long-range) degrees of freedom are included
explicitly, with the rest parametrized in terms of the most general
local (contact) interactions. This procedure exploits the fact that
a low-energy probe of momentum $k$ cannot resolve structures on 
scales smaller than $R\sim 1/k$. (Note that $\hbar=c=1$ in this talk.)
Using renormalization,
the influence of short-distance physics on low-energy observables
is captured in a small number of low-energy constants.
Thus, the EFT describes universal low-energy physics independent of
detailed assumptions about the short-distance dynamics. 
All physical observables can be described in a controlled expansion in 
powers of $kl$, where $l$ is the characteristic low-energy 
length scale of the system. The size of $l$ depends on the system 
under consideration: for a finite range potential, e.g., it is given 
by the range of the potential.
For the systems discussed here, $l$ is of the order of the effective 
range $r_e$.

In this talk, I will focus on applications of EFT to few-body systems
with large S-wave scattering length $a \gg l$. 
For a generic system, the scattering length
is of the same order of magnitude as the low-energy length scale $l$.
Only a very specific choice of the parameters in the underlying theory 
(a so-called {\it fine tuning}) will generate a large scattering length.
Systems with large scattering length can be found in many
areas of physics. Examples are the S-wave scattering of nucleons
and of $^4$He atoms. For alkali atoms close to a Feshbach resonance,
$a$ can be tuned experimentally by adjusting an external magnetic field.

\section{THREE-BODY SYSTEM WITH LARGE SCATTERING LENGTH}

In this section, we give a very brief review of the EFT
for few-body systems with large scattering length $a$. 
We will focus on S-waves (For more details, see Ref.~\cite{BhvK99}).

For typical momenta $k\sim 1/a$,
the EFT expansion is in powers of $r_e/a$ so that higher
order corrections  are suppressed by powers of $r_e/a$. The leading order
corresponds to $r_e=0$.
We consider a 2-body system of nonrelativistic bosonic atoms 
with large scattering length $a$ and mass $m$.
At sufficiently low energies, the most general Lagrangian 
for S-wave interactions may be written as:
\begin{equation}
{\cal L}  =  \psi^\dagger
             \left(i\partial_{t}+\frac{\vec{\nabla}^{2}}{2m}\right)\psi
 - \frac{C_0}{2} (\psi^\dagger \psi)^2
 - \frac{D_0}{6} (\psi^\dagger\psi)^3 + \ldots\,,
\label{eq:eftlag}
\end{equation}
where the $C_0$ and $D_0$ are nonderivative 2- and 3-body interaction
terms, respectively. The strength of the $C_0$ term is determined by
the scattering length $a$, while $D_0$ depends on a 3-body parameter
to be introduced below.
The dots represent higher-order derivative terms
which are suppressed at low-energies. For momenta $k$ of the order of
the inverse scattering length $1/a$, the problem is nonperturbative 
in $ka$. The exact 2-body scattering amplitude can be obtained
analytically by summing the so-called {\it bubble diagrams} with the
$C_0$ interaction term. The $D_0$ term does not contribute to 2-body
observables. After renormalization, 
the resulting amplitude reproduces the leading order of the 
well-known effective range expansion for the atom-atom
scattering amplitude:~$f_{AA}(k)=(-1/a -ik)^{-1}\,,$
where the total energy is $E=k^2/m$. 
If $a>0$, $f_{AA}$ has a pole at $k=i/a$ corresponding
to a shallow dimer with binding energy $B_2=1/(ma^2)$. 
Higher-order derivative
interactions are perturbative and give the momentum-dependent terms
in the effective range expansion.

\begin{figure}[htb]
\centerline{\includegraphics*[width=14cm,angle=0]{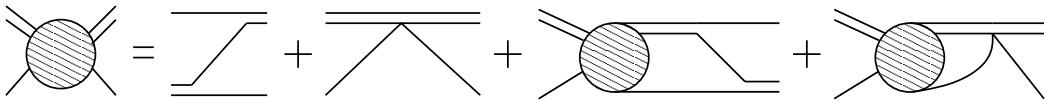}}
\caption{Integral equation for the atom-dimer scattering amplitude.
Single (double) line indicates single atom (two-atom state).}
\label{fig:ineq}
\end{figure}

We now turn to the 3-body system. Here, it is useful to introduce
an auxiliary field for the two-atom state (see Ref.~\cite{BhvK99}
for details).
At leading order, the atom-dimer scattering amplitude is given by the 
integral equation shown in  Fig.~\ref{fig:ineq}. 
A solid line indicates a single atom and a double line indicates an 
interacting two-atom state (including rescattering corrections).
The integral equation contains contributions from both
the 2-body and the 3-body force terms.
The inhomogeneous term is given by the first two diagrams on the 
right-hand side: the one-atom exchange diagram
and the 3-body force. The integral equation simply sums
these diagrams to all orders.
After projecting onto S-waves, we obtain the equation
\begin{eqnarray}
{\cal T} (k, p; E) & = & {16 \over 3 a}\, M(k,p;E)
+ {4 \over \pi} \int_0^\Lambda 
{dq \, q^2 \, M(q,p;E)\over  -{1/a} + \sqrt{3q^2/4 -mE
-i \epsilon}}\, {\cal T} (k, q; E)\,,
\label{eq:BHvK}
\end{eqnarray}
for the off-shell atom-dimer scattering amplitude with the inhomogeneous
term 
\begin{eqnarray}
M(k,p;E)&=& {1 \over 2pk} \ln \left({p^2 + pk + k^2 -mE \over
p^2 - pk + k^2 - mE}\right) + {H(\Lambda) \over \Lambda^2}\,.
\end{eqnarray}
The logarithmic term is the S-wave projected one-atom exchange,
while the term proportional to $H(\Lambda)$ comes from the 3-body
force. The physical atom-dimer scattering amplitude 
$f_{AD}$ is given by the solution ${\cal T}$ evaluated at the on-shell 
point:~$f_{AD}(k) = {\cal T} (k, k; E)$ where
$E= 3k^2/(4m)-1/(ma^2)\,$.
The 3-body binding energies $B_3$ are given by those values of $E$
for which the homogeneous version of Eq.~(\ref{eq:BHvK}) has a
nontrivial solution.

Note that an ultraviolet cutoff $\Lambda$ has been introduced in
(\ref{eq:BHvK}). This cutoff is required to insure that Eq.~(\ref{eq:BHvK})
has a unique solution. All physical observables, however, must be invariant
under changes of the cutoff, which determines the behavior of $H$ as a 
function of $\Lambda$ \cite{BhvK99}:
\begin{eqnarray}
H (\Lambda) = {\cos [s_0 \ln (\Lambda/ \Lambda_*) + \arctan s_0]
\over \cos [s_0 \ln (\Lambda/ \Lambda_*) - \arctan s_0]}\,,
\label{H-Lambda}
\end{eqnarray}
where $s_0=1.00624$ is a transcendental number and $\Lambda_*$
is a 3-body parameter introduced by dimensional transmutation. 
This parameter cannot be predicted by the EFT and must be taken from 
experiment. Note that $H (\Lambda)$ is periodic and runs on
a limit cycle. When $\Lambda$ is increased by a factor of
$\exp(\pi/s_0)\approx 22.7$, $H (\Lambda)$ returns to its original 
value. 

In summary, two parameters are required in the 
3-body system at leading order in $r_e/a$:
the scattering length $a$ (or the dimer
binding energy $B_2$) and the 3-body parameter $\Lambda_*$ \cite{BhvK99}. 
The EFT reproduces the universal aspects of the 3-body system that
were first derived by Efimov \cite{Efi71}. These include the 
accumulation of infinitely many 3-body bound states 
(so-called {\it Efimov states}) at threshold as $a\to\pm\infty$.
As we will demonstrate in the next section, EFT is a very efficient
calculational tool to calculate those properties.

\section{APPLICATION TO COLD ATOMS AND BEC}

In this section, we discuss some applications of the EFT to systems of cold
atoms and BEC's.

First we turn to the 3-body system of $^4$He atoms, where
$a/r_e \approx 15$. Both the $^4$He dimer and trimer were observed.
The complete 3-body bound state spectrum in leading order in the EFT 
follows from solving Eq.~(\ref{eq:BHvK}). It can also be 
parametrized in terms of a universal function $\Delta$ of the
angle $\xi=a\sqrt{mB_3}$ and a 3-body parameter \cite{Efi71}.
The function $\Delta(\xi)$ was recently calculated in EFT
\cite{univ}.
Unfortunately, no 3-body observables have been measured for $^4$He
atoms up to now, so that $\Lambda_*$ can not be determined from experiment.
However, extensive bound state calculations with modern potentials 
exist.
We have used recent calculations by Motovilov et al.~\cite{Moto}, to obtain 
the dimer binding energy and the 3-body parameter $\Lambda_*$ 
from the calculated binding energy of the trimer excited state $B_3^{(1)}$.
We have then used this input to calculate the trimer ground state
energy  $B_3^{(0)}$ for four different $^4$He potentials 
\cite{he4univ}. The results are shown in Table \ref{tab:4he}.
\begin{table}[htb]
\caption{Universality for $^4$He atoms. The first three columns show the
calculated binding energies of the dimer and the trimer ground and excited 
states for four modern $^4$He potentials \protect\cite{Moto}. 
The last two columns show the extracted value of $a\Lambda_*$ 
and the leading order EFT prediction for the trimer ground state
\cite{he4univ}.}
\label{tab:4he}
\begin{tabular}{c||c|c|c||c|c}
Potential & $B_2$ & $B_3^{(0)}$ & $B_3^{(1)}$ &
$a\Lambda_*$ & $B_3^{(0)}$(pred.) \\
\hline\hline
HFDHE2 & 0.830 & 116.7 & 1.67 & 1.258 & 118\\
HFD-B  & 1.685 & 132.5 & 2.74 & 0.922 & 138\\
LM2M2  & 1.303 & 125.9 & 2.28 & 1.033 & 130\\
TTY    & 1.310 & 125.8 & 2.28 & 1.025 & 129
\end{tabular}
\end{table}
The predicted ground state binding energy $B_3^{(0)}$
agrees well with the direct calculation of Motovilov et al.~\cite{Moto}.
This demonstrates that both the excited and ground state of the $^4$He
trimer are Efimov states.

Universality also constrains 3-body scattering observables. For example
the atom-dimer scattering length can be expressed in terms of $a$
and $\Lambda_*$ as \cite{BhvK99,Efi71}
\begin{equation}
a_{AD}=a\, (1.46-2.15\tan [s_0 \ln(a\Lambda_*) +0.09])\,
(1\,+\,{\cal O}(r_e/a))\,,\qquad\qquad a>0\,.
\end{equation}
where the numerical constants were calculated in EFT. 
Using the value of $\Lambda_*$ extracted from the 
excited state binding energy, a good agreement with $a_{AD}$ from the 
direct calculation of Ref.~\cite{Moto} is obtained \cite{he4univ}. 

\begin{figure}[htb]
\centerline{\includegraphics*[width=8.8cm,angle=0]{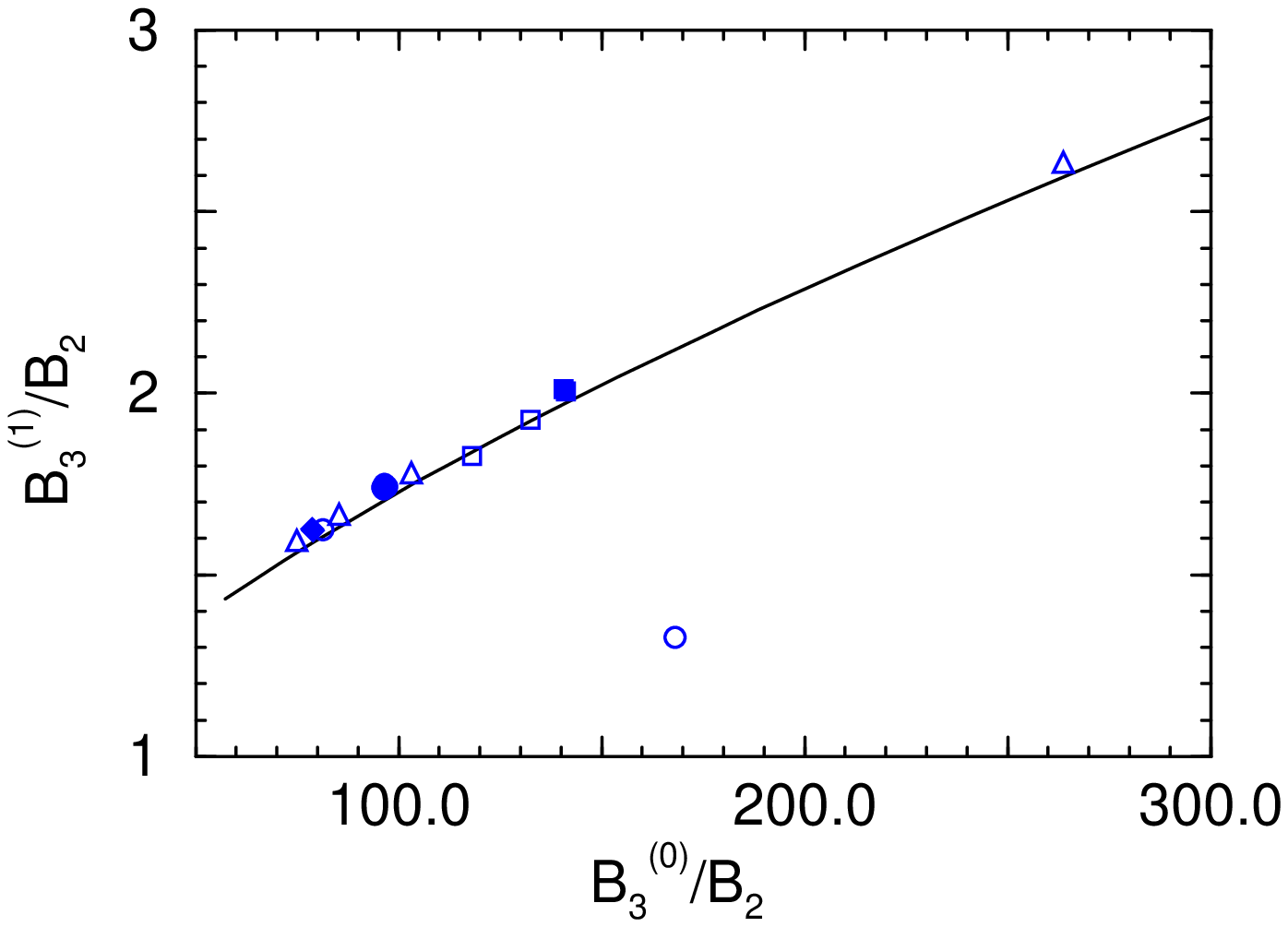}
\includegraphics*[width=8.2cm,angle=0]{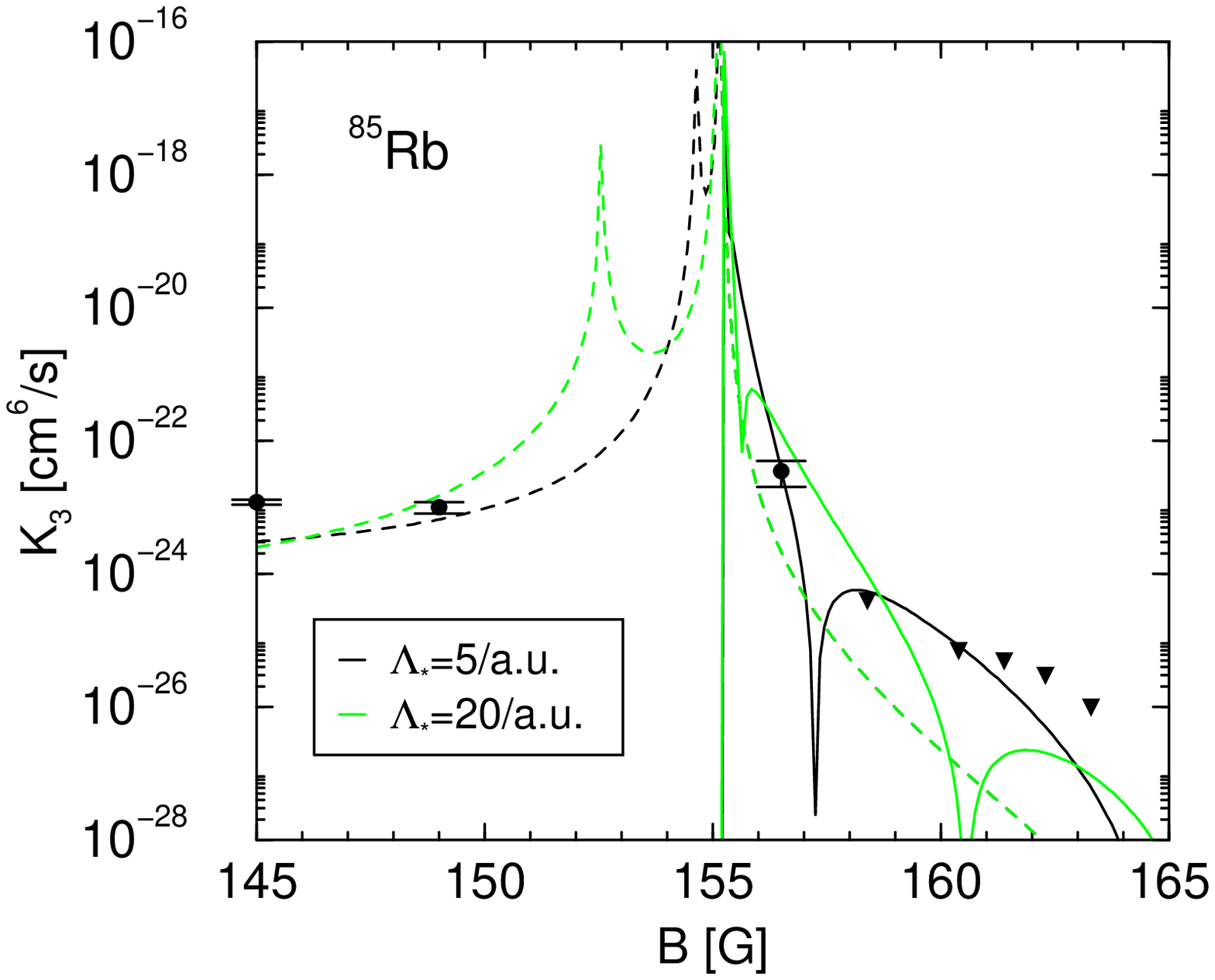}}
\caption{Left panel: The scaling function $B_3^{(1)}/B_2$ vs. 
$B_3^{(0)}/B_2$ for $^4$He atoms. Right panel: The loss rate coefficient 
$K_3$ in $^{85}$Rb as a function of the external magnetic field
for different values of $\Lambda_*$. Solid (dashed) line gives 
contributions of shallow (deep) bound states. Data points are from
Ref.~\protect\cite{jila}.}
\label{fig:scale_K3}
\end{figure}

Universality is also manifest in universal scaling functions. 
In the left panel of Fig.~\ref{fig:scale_K3}, we display the
scaling function relating $B_3^{(1)}/B_2$ to  $B_3^{(0)}/B_2$ \cite{he4univ}.
The data points give various calculations using modern $^4$He potentials 
while the solid line gives the universality prediction from EFT. 
Different points on this line correspond to different values of
$\Lambda_*$. The small deviations of the 
potential calculations from the universal curve are mainly due to 
effective range corrections and can be calculated at next-to-leading
order in EFT. The calculation corresponding to the
data point far off the universal curve can easily be 
identified as problematic.

Next we turn to 3-body recombination, which is the process when
three atoms scatter to form a dimer and the third atom balances energy
and momentum. This is one of the main loss processes for trapped atoms
and condensates of atoms near a Feshbach resonance. The event rate 
can be parametrized as $\nu=\alpha\rho^3$, where $\rho$ is the density
of the atoms and $\alpha$ is the recombination constant. 
At threshold, the leading order EFT result for the 
contribution to $\alpha$ from recombination into the shallow dimer 
is \cite{3brec}
\begin{equation}
\alpha=67.1 \,\frac{a^4}{m}\,\sin^2 [s_0 \ln (a\Lambda_*)+0.19]\,
(1\,+\,{\cal O}(r_e/a))\,,\qquad\qquad a>0\,.
\end{equation}
If deeply bound dimers are present, there are additional contributions
from recombination into the deep bound states. If $a<0$, 
recombination can only go into the deep states.

In the right panel of Fig.~\ref{fig:scale_K3}, we show the
3-body loss rate coefficient $K_3=3\alpha$ from Ref.~\cite{3brec}
for two different values of
$\Lambda_*$  compared to experimental data
for 3-body losses in a cold gas of $^{85}$Rb atoms near a
Feshbach resonance at $B=155$ G \cite{jila}. 
The value $\Lambda_*=5$/a.u. seems to be preferred by the data.

\section{SUMMARY \& OUTLOOK}

The renormalization of the EFT for 3-body systems with large scattering 
length requires a 3-body force at leading order \cite{BhvK99}. 
The renormalization 
group evolution of the 3-body force is governed by a limit cycle.
Via dimensional transmutation, the 3-body force introduces a dimensionful 
parameter, $\Lambda_*$, that parametrizes universal
relations between different 3-body observables.

The EFT is very general and can be applied to many
physical systems ranging from the 3-body system of $^4$He atoms
\cite{he4univ}, to 3-body recombination and dimer deactivation
in cold atomic gases/BEC's \cite{3brec,dimdeac}, 
to the triton and hypertriton in nuclear physics \cite{trithypt}.

Future challenges include the extension of the EFT to the four-body system,
the stability and phase structure of cold atomic gases/BEC's close to
a Feshbach resonance \cite{Braaten:2001ay},
and the possibility of coexisting condensates of atoms,
dimers, and trimers \cite{Braaten:2002er}.

\end{document}